\documentclass[10pt,journal]{IEEEtran}
\usepackage{amsmath}
\usepackage{graphicx}
\usepackage{color}
\begin{document}
\title{Physical Layer Security Schemes for Full-Duplex Cooperative Systems: State of the Art and Beyond}
\author{Binh Van Nguyen, Hyoyoung Jung, and Kiseon Kim}
\maketitle
\begin{abstract}
Due to the broadcast nature of wireless medium, wireless communication is highly vulnerable to eavesdropping attack. Traditionally, secure wireless data transmission has relied on cryptographic techniques at the network layer which incur high computational power and complexity. As an alternative, physical layer security (PLS) is emerging as a promising paradigm to protect wireless systems by exploiting the physical characteristics of the wireless channels. Among various PLS approaches, the one based on cooperative communication is favorable and has got a lot of interest from the research community. Although PLS schemes with half-duplex relays have been extensively discovered, the issue of PLS in cooperative systems with full-duplex (FD) relays is far from being comprehensively understood. In this paper, we first present the state of the art on PLS approaches proposed for FD cooperative systems. We then provide a case study in which a source-based jamming scheme is proposed to enhance the secrecy performance of a cooperative system with an untrusted FD relay. Finally, we outline several interesting yet challenging future research problems in this topic.
\end{abstract}
\IEEEpeerreviewmaketitle
\section{Introduction}
Owning to the recent evolution of wireless communications and the popularity of hand-held devices such as smart phones and tablets, more and more people are using wireless networks for e-banking, personal emails, e-health, and file sharing. However, due to the broadcast nature of wireless medium, confidential information can be easily overheard by adversaries. It has recently reported that an increasing number of wireless devices are abused for malicious attacks, data forging, financial information theft, online bullying, and so on \cite{Zou'16}. Therefore, ensuring secrecy and privacy are of utmost concern for future wireless communication systems. Traditionally, secure wireless data transmission has been relied on the cryptographic technique at the network layer which incurs a very high computational power and complexity. As an alternative, physical layer security (PLS), or information-theoretic security, is emerging as a promising paradigm to realize secure communication against eavesdropping attacks by exploiting the characteristics of wireless channels \cite{Mukherjee'14}. The advantages of PLS lie behind the fact that it is simple, it does not require any assumption at adversaries, and it can be used to augment existing cryptographic schemes, i.e. physical layer noise was reported to considerably improve the performance of cryptographic schemes in \cite{Oggier'14} and references therein.

On the other hand, cooperative communication has been proven to be a powerful method to increase the throughput and coverage of single-antenna systems \cite{Cuba'12}. In a cooperative system, single or multiple neighbor nodes of a source, called relays, help the source to forward manipulated versions of the source signal to a destination. The way a relay manipulates the source signal depends on which relaying protocol is deployed. Among various relaying protocols, the most popular are amplify-and-forward (AnF) and decode-and-forward (DnF). An AnF relay simply amplifies its received signal and forwards the outcome to the destination. In addition, a DnF relay first decodes its received signal, re-encodes, and then forwards the result to the destination. Besides relaying protocols, the operation of a relay also depends on relaying modes which include half-duplex (HD) and full-duplex (FD) modes. In the HD mode, relays receive and forward the source signal in different time slots. On the contrary, in the FD mode, relays can receive and transmit simultaneously. As a result, systems with FD relays, referred to as FD systems, have a better spectral efficiency than that given by the HD counterparts. The FD mode was considered impractical in the past since the performance of FD systems is limited by the loop interference (LI) between a relay input and output. Recently, with advances on antenna design and signal processing the LI can be significantly canceled, and thus, the FD mode has got a lot of attention from research community \cite{Zhang'15}.

It is shown in \cite{Dong'10} that cooperative communication also provides a great potential to secure wireless data transmissions, which provokes a significant research interest in the topic of designing PLS schemes based on cooperative nodes, i.e. [1]-[2] and references therein. PLS for cooperative systems can be categorized as secret key-based and keyless schemes, among which keyless PLS (K-PLS) schemes have been largely considered because they do not require secret keys for encryption/decryption data \cite{Zou'16}. Keyless PLS schemes for HD systems have been extensively investigated in \cite{Rodriguez'15}, \cite{Wang'15}, and references therein. It has been shown that when relays are trusted, they can be exploited to enhance the systems secrecy performance by using the relaying, jamming, and the hybrid relaying and jamming approaches. In addition, even when relays are untrustworthy, it is still possible to obtain secure communication by applying the destination-based or the source-destination-based jamming schemes. Although K-PLS schemes for HD systems are well explored, research works on FD systems are limited \cite{Chen'15}-\cite{Zhu'16}. It means that designing and analyzing K-PLS schemes for FD systems are only at their early stages, and the opportunity for innovation remains tremendous. This observation motivates us to present a review on the current state of the art in this line of research and discuss possible future research directions. Note that we only focus on the PLS of FD cooperative systems rather than providing a comprehensive survey in the whole field of cooperative PLS.

\section{Fundamentals of Physical Layer Security in Cooperative Systems}
Before going into details about our main focus, we first quickly present the fundamental concepts of PLS to make our paper easy to follow by general readers. The basic principle of PLS is to exploit the physical characteristics of the wireless medium, i.e. fading, noise, and interferences, so as to limit the amount of information that can be extracted at the bit level by eavesdroppers. The main advantages of PLS come from the facts that no computational restrictions are placed on the eavesdroppers, PLS can operate independently of higher layers, and that very precise statements can be made about the information that is leaked to the eavesdroppers as a function of the channel quality \cite{Mukherjee'14}. Apparently, the aforementioned advantages of PLS is only half of the story. We should also note that PLS relies on average information measures. A system can be designed for a specific level of security, claiming for instance that a block will be secure with a very high probability; however, it might not be possible to guarantee confidentiality with probability one.

The research on PLS was initiated by Wyner in \cite{Wyner'75} where a three-node configuration including a source, a destination, and an eavesdropper (as shown in the Fig. 1) was considered. It was shown that secure data transmission can be achieved from an information theoretic perspective if the source-eavesdropper channel, referred to as the wiretap channel, is a degraded version of the legitimate source-destination channel. In other words, the system can realize secure communication if its secrecy capacity (SC), which is defined as the difference between the capacity obtained at the legitimate receiver and that obtained at the eavesdropper, is positive. The SC represents the maximum transmission rate at which the source can communicate with the destination without the eavesdropper being able to acquire any confidential information. If the destination imposes a target secrecy rate higher than the SC, a secrecy outage event will occur. In addition, the average probability of this event is called secrecy outage probability (SOP), which is considered as the most general secrecy performance measure of a system under PLS constraint.

After Wyner's seminal work, considerable research efforts have been devoted to develop various PLS techniques which can be generally classified into artificial-noise aided, multi-antenna diversity, cooperative diversity, secret key generation, and coding \cite{Zou'16}-\cite{Mukherjee'14}, among which the one based on cooperative communication is of our special interest. PLS for cooperative systems can be further categorized as secret key-based and keyless schemes \cite{Zou'16}. In a key-based scheme, a source and a destination first transmit known signals to each other via a relay. The two nodes then estimate a virtual channel between them from the received signals. Thereafter, several necessary processes are carried out to make sure that keys generated from the virtual channel on both sides are the same. Finally, this key is used to encrypt and decrypt confidential messages. Although a relay can be used to improve the key generation rate, it can be compromised and become a malicious user. Consequently, secret information can be easily intercepted and revealed, which may be the main reason why research works on key-based PLS schemes for cooperative systems are very limited \cite{Zou'16}. On the other hand, K-PLS schemes, which do not require secret keys for encryption/decryption data but employ signal processing and diversity techniques to obtain secure transmissions, have got a lot of interest from research community. The ultimate goal of a K-PLS scheme is to improve the SC of a considered system either by enhancing the capacity obtained at the receiver or by degrading the capacity achieved at the eavesdropper.
\begin{figure}[t]
  \centering
  \includegraphics[width=7.5cm]{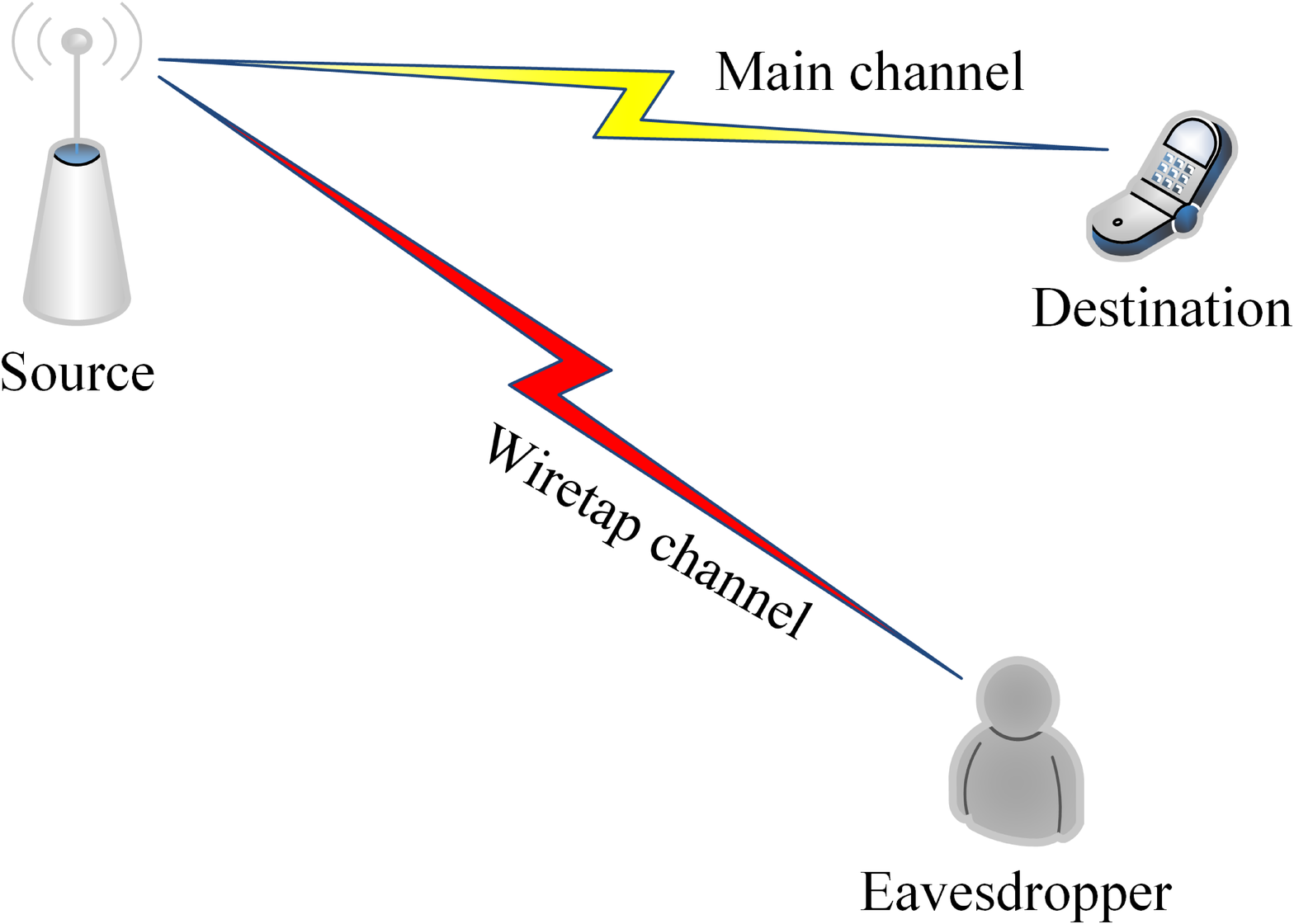}
  \caption{Wyner's wiretap channel model.}
\end{figure}

\section{Keyless PLS Schemes for FD Systems with Trusted Relays}
\begin{table*}[!t]
  \centering
  \caption{A Summary of PLS Schemes Proposed for Cooperative Systems with Full-Duplex Relays}
  \includegraphics[width = 16cm]{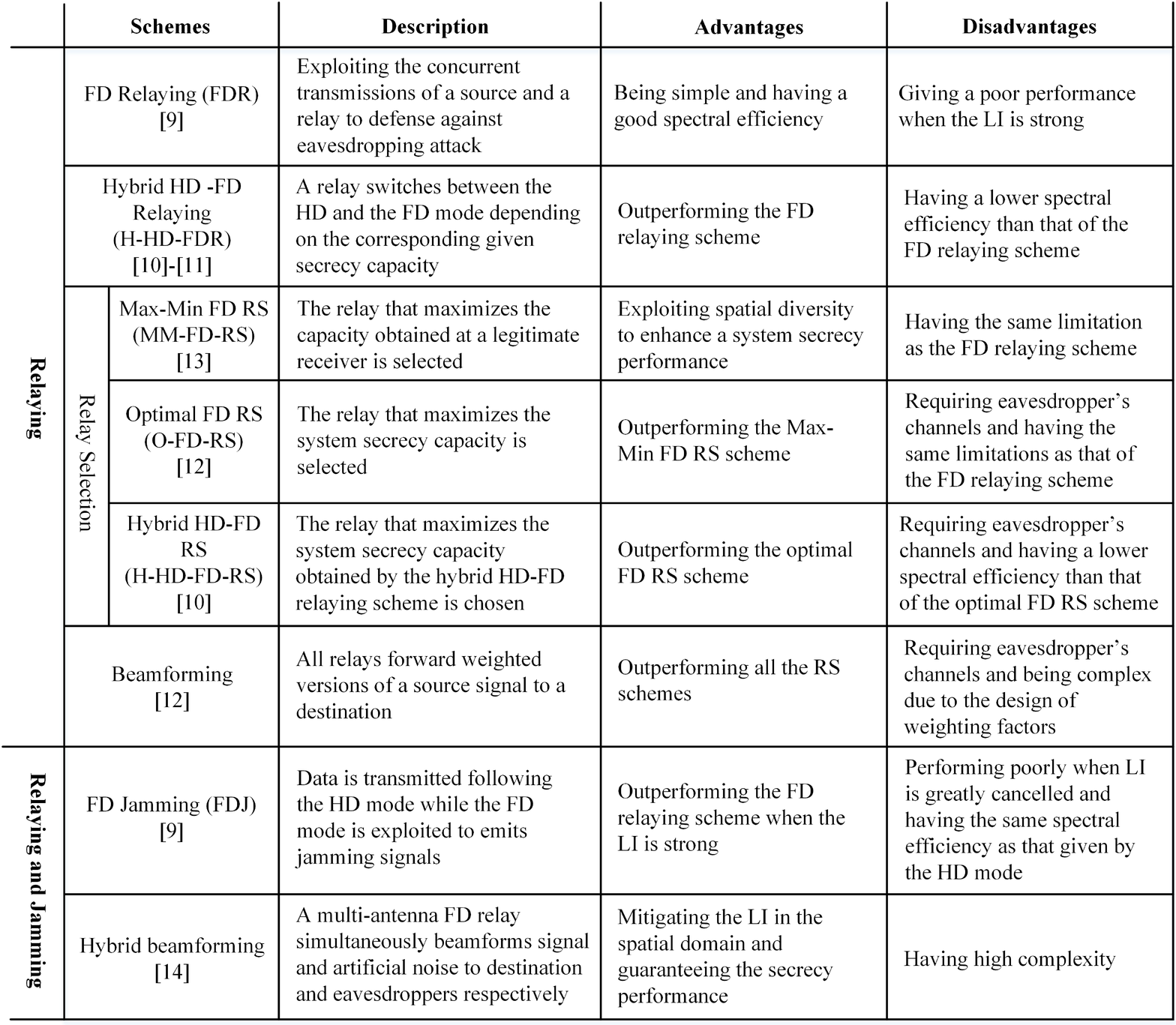} \\
  \vspace*{2pt}
  \hrulefill
\end{table*}
Although FD relays have been largely considered for conventional cooperative systems (systems without an eavesdropper), analyzing the capability of FD relays for the security purpose has only carried out recently in \cite{Chen'15}. K-PLS schemes for FD systems can be categorized into different groups based upon the role of relays. Particularly, we have relaying group, in which relay(s) only play(s) the role of pure (actual) relay(s), and relaying and jamming group, in which relay(s) play(s) the role of both actual relay(s) and friendly jammer(s). In addition, in the former group, schemes with a single relay consist of FD relaying (FDR) and hybrid HD-FD relaying (H-HD-FDR), while schemes with multiple relays include beamforming and relay selection (RS). Moreover, the latter group contains the FD jamming (FDJ) scheme with one relay and the hybrid beamforming scheme with multiple relays. A summary of K-PLS schemes for FD systems is given in the Table I, shown at the top of the next page.

\subsection{Relaying Schemes}
When only one relay is available, the relay can be operated following the FDR or the H-HD-FDR schemes \cite{Chen'15}-\cite{Shafie'16}. In the FDR scheme, the relay simply operates in the FD mode, i.e. simultaneously receiving and forwarding the source messages. Although an eavesdropper can receive the signals coming from the source and the relay at the same time, the relay transmission signal is a delayed version of the source transmission signal. Consequently, the concurrent transmissions of the source and the relay cause inter-symbol interference at the eavesdropper, from which the achievable capacity obtained at the eavesdropper is degraded. It is shown in \cite{Chen'15} that when the LI between the relay input and output is sufficiently canceled, which is feasible nowadays thanks to advances in antenna design and signal processing, the FDR scheme can provide a much lower SOP than that given by the HD relaying (HDR) counterpart. However, when the residual LI (after LI cancellation) is strong, the converse holds. It means that between the HD and the FD mode, one of them can be superior to the other depending upon the level of the residual LI. Motivated by this observation, the authors of \cite{He'16} propose the H-HD-FDR scheme, in which the relay switches between the two modes to achieve the best secrecy performance. More specifically, if the system SC obtained by the HD mode is larger than that given by the FD mode, the relay will operate following the HD fashion, otherwise; the FD mode will be employed. It is shown that the hybrid scheme provides a good secrecy performance compared to that of the FDR and the HDR schemes. The secrecy performance of the H-HD-FDR scheme can be further boosted by employing a buffer at the relay \cite{Shafie'16}. Since received packets can be stored into the buffer, the relay can dynamically switch between the source-relay and the relay-destination channels (under the HD mode), and thus, the system secrecy performance can be significantly increased.

On the other hand, when multiple relays are available, the best scheme is beamforming \cite{Lee'16}. In this scheme, multiple relays perform beamforming to cancel the confidential information at the eavesdropper. Particularly, the zero-forcing technique is incorporated with the max-min fair beamforming approach to derive the beamforming weighting factors. If the channel state information (CSI) of all links including the wiretap channels is available, completely nulling out confidential signals at the eavesdropper can be obtained. Although the beamforming scheme can provide a very promising secrecy performance, its deployment complexity is high due to the requirement of CSI of the wiretap channels and the difficulties related to designing the weighting factors. Alternatives to the beamforming scheme are RS schems, which include Max-Min FD RS (MM-FD-RS), optimal FD RS (O-FD-RS), and hybrid HD-FD RS (H-HD-FD-RS) schemes \cite{Nguyen'15}, \cite{He'16}. In the MM-FD-RS/O-FD-RS scheme, all the relays operate in the FD mode and the relay that maximizes the conventional capacity/SC is selected. In addition, in the H-HD-FD-RS scheme, each relay first selects its best relaying mode, i.e. HD or FD, and then the relay that maximizes the system SC is chosen. It is shown that the H-HD-FD-RS scheme outperforms the O-FD-RS counterpart, which is also superior to the MM-FD-RS scheme. However, it should be noted that the MM-FD-RS scheme is the simplest one since it does not require CSI of the wiretap channels which is quite challenging to be obtained in reality.

\subsection{Relaying and Jamming Schemes}
Different from the aforementioned schemes, the authors of \cite{Chen'15} propose the FDJ scheme in which a relay plays the role of an actual relay or a friendly jammer one after the other, as illustrated in Fig. 2. In particular, in this scheme, each transmission takes place in two consecutive phases. In the first phase, while receiving the source information, the relay simultaneously transmits jamming signals to confuse the eavesdropper. In addition, in the second phase, while the relay forwards the confidential information to the destination, the source emits jamming signals to jam the eavesdropper. We can see that the eavesdropper always receives data from one node and intended jamming signals from another node. It is shown that when the target secrecy rate is small, the FDJ scheme outperforms the FDR counterpart. However, when the target secrecy rate becomes larger, the converse holds.
\begin{figure}[t]
  \centering
  \includegraphics[width=6.5cm]{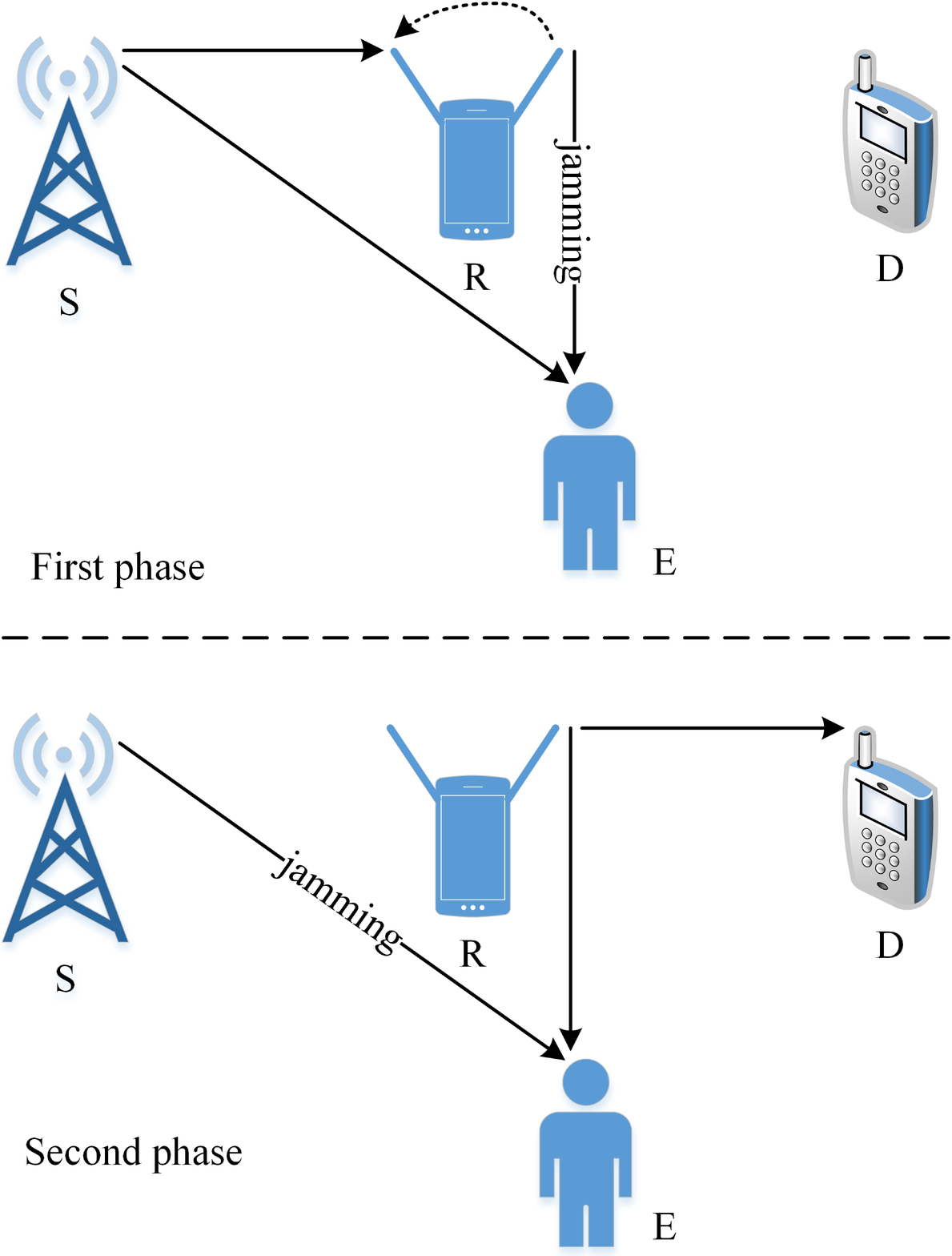}
  \caption{An illustration of the FDJ scheme, the FD capability of the relay is exploited to transmit jamming signals toward the eavesdropper.}
\end{figure}

When the FD relay can be equipped with multiple receive and transmit antennas, the authors of \cite{Zhu'16} propose the hybrid beamforming scheme, in which the relay simultaneously transmits information to the destination and artificial noise to the eavesdropper. The information and the artificial noise beamforming vectors are designed to minimize the relay's transmit power while guaranteeing the system secrecy performance. The proposed scheme is also able to mitigate the LI in the spatial domain, and thus, digital interference cancellation circuits can be omitted. However, equipping multiple antennas at a relay node is not always feasible due to size, cost, or hardware limitations. In addition, the scheme incurs a high computation complexity, which limit its applications.

\subsection{Secrecy Performance Comparison}
In this subsection, we shall present comparisons of the secrecy performance of the aforementioned K-PLS schemes with single relay and with multiple relays, respectively. The SOPs of the existing K-PLS schemes with single relay versus the average SNR of the residual LI channel $\gamma_{rr}$ are illustrated in the Fig. 3, shown at the top of the next page. Thereby, the average SNRs of the source-relay, relay-destination, source-eavesdropper, and the relay-eavesdropper are set to be $40$, $40$, $10$, and $10$ dB. The independent Rayleigh fading is also assumed. As expected, the FDR scheme outperforms the HD counterpart when the residual LI is small, however, the converse holds otherwise. In addition, it is shown that the secrecy performance of the H-HD-FD scheme converges to that of the FDR and HDR schemes in the low and the high region of $\gamma_{rr}$, respectively. And in the medium range of $\gamma_{rr}$ the H-HD-FD scheme provides a better secrecy performance than that given by the FDR and the HDR counterparts. Moreover, it is observed that the FDJ scheme has a better secrecy performance than that of the FDR scheme when the target secrecy rate $R_0 = 1$. However, the FDJ is inferior to the FDR counterpart when $R_0 = 2$. The reason is that the data transmission in the FDJ scheme is operated following the HD manner. Consequently, there is a factor of $1/2$ in the secrecy capacity of the FDJ scheme which deteriorates the system secrecy performance. When $R_0$ is small, the effect of the intended jamming signals is larger than that of the $1/2$ factor, and thus, the FDJ scheme gives a better secrecy performance than that provided by the FDR scheme. However, when $R_0$ becomes higher, the effect of the $1/2$ factor goes up and becomes dominant. Consequently, the converse happens. Lastly, the results presented in the Fig. 3 suggest that combining the H-HD-FD and the FDJ schemes is a promising solution to enhance the secrecy performance of FD cooperative systems in the low regions of $\gamma_{rr}$ and $R_0$.
\begin{figure}[t]
  \centering
  \includegraphics[width=8.5cm]{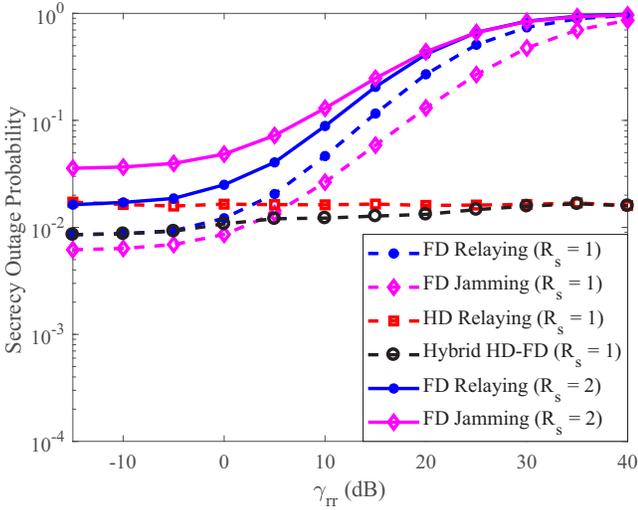}
  \caption{SOPs of the existing K-PLS schemes with a single relay, $\gamma_{sr} = \gamma_{rd} = 40$ dB and $\gamma_{se} = \gamma_{re} = 10$ dB.}
\end{figure}

We present the SOPs of the O-FD-RS, MM-FD-RS, H-HD-FD-RS, and the beamforming schemes versus $\gamma_{rr}$ in the Fig. 4, shown at the top of the next page. For a comparison purpose, the SOPs of the optimal HD RS and the naive random FD RS schemes are also plotted. In the optimal HD RS scheme, all relays operate in the HD mode and the relay maximizing the system SC will be selected. In addition, in the random FD RS scheme, all relays deploy the FD mode and the active relay is randomly chosen. Here, the simulation settings are as follows: the number of relays is 4 and the average SNRs of the source-relay, relay-destination, source-eavesdropper, and the relay-eavesdropper are set to be $30$, $30$, $10$, $10$ dB, respectively. In addition, for the beamforming scheme, the inter-relay interferences are not considered since the relays can be far away located, and thus, the inter-relay interferences can be canceled much easier than the LI. It is first shown that the random FD RS scheme gives the worst SOP over the whole range of $\gamma_{rr}$, while the beamforming scheme provides the best secrecy performance in the low region of $\gamma_{rr}$. On the other hand, the optimal HD RS and the H-HD-FD-RS schemes produce the best secrecy performance in the high region of $\gamma_{rr}$. This fact is understandable because the schemes solely based on the FD mode only work well when the residual LI is weak, while the optimal HD RS and the H-HD-FD-RS schemes are based on the HD mode which are well-known to be superior to the FD counterpart when the residual LI is strong. Secondly, the figure also shows that the MM-FD-RS scheme gives a comparable secrecy performance with that of the O-FD-RS scheme, which possesses a much higher deployment complexity due to the requirement of CSI of the eavesdropping channels. Thirdly, the results given in the Fig. 4 recommend that in the high region of $\gamma_{rr}$, the optimal HD RS scheme should be used. In addition, in the low region of $\gamma_{rr}$, although the O-FD-RS, H-HD-FD-RS, and the beamforming schemes provide excellent SOPs, they incur high complexity, and thus the MM-FD-RS scheme turns out to be a good alternative. In summary, replacing the O-FD-RS criterion in the H-HD-FD-RS scheme by the MM-FD-RS criterion to generate a novel modified H-HD-FD-RS scheme is a low cost yet promising solution to guarantee the system secrecy performance.
\begin{figure}[t]
  \centering
  \includegraphics[width=8.5cm]{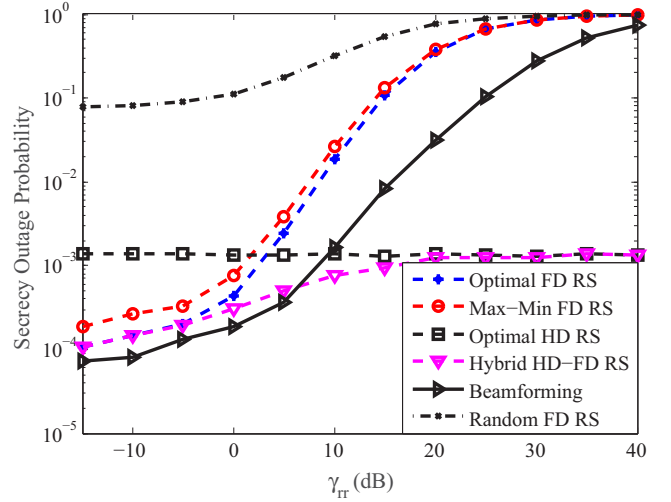}
  \caption{SOPs of the relay selection and the beamforming schemes with $\gamma_{sr} = \gamma_{rd} = 30$ dB, $\gamma_{se} = \gamma_{re} = 10$ dB, and $4$ relays.}
\end{figure}

\section{A Case of Study: A Source-based Jamming Scheme for a System with an Untrusted FD Relay}
We observe that all aforementioned works only focus on trusted FD relays, while ignoring the scenarios with untrustworthy relays. The issue of untrusted relays arises in systems in which nodes have different level of authority. For example, in ad-hoc systems, relays are needed for connectivity; however, they can be unauthenticated. As an another example, in government or financial institution systems, relays may have a lower security clearance than that of the source-destination pair. In such systems, relays assist the source-destination communication, yet they simultaneously try to eavesdrop confidential information. Although cooperative systems with untrusted FD relays are practical, they have not been investigated yet. Motivated by this observation, we now consider a cooperative system with an untrusted FD relay, which is graphical illustrated in the Fig. 5. Particularly, the considered system consists of  a single-antenna source $S$, a single-antenna destination $D$, and an untrusted FD relay $R$. The relay has two antennas used for reception and transmission, respectively. In addition, $R$ employs the AnF relaying protocol which is obviously more preferable than the DnF counterpart under untrusted relay's scenarios.
\begin{figure}[t]
  \centering
  \includegraphics[width=7cm]{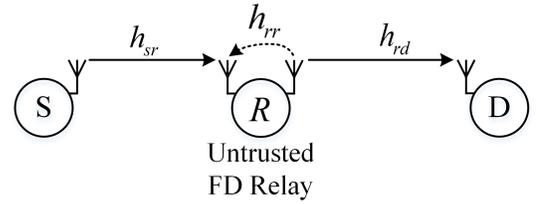}
  \caption{A wireless cooperative system with an untrusted full-duplex relay.}
\end{figure}

It can be readily verified that the capacity obtained at $D$  is always less than or equal to that obtained at $R$. In other words, the system secrecy capacity is always zero, and thus, secure data transmission cannot be achieved. To enhance the system's secrecy performance, we propose to use a source-based jamming (SBJ) scheme, in which the source uses a fraction of its power to emit jamming signal to degrade the relay's interception. More specifically, at a time instant $t$, $S$ transmits a composite signal containing the confidential signal $x_s$ with power $\alpha P_s$ and the jamming signal $x_j$ with power $\left( {1 - \alpha } \right){P_s}$ to $R$, where $P_s$ and $0 \le \alpha  \le 1$ denote the source's transmit power and the power allocation ratio between $x_s$ and $x_j$. The relay, while receiving the source's information, transmits an amplified version of of its received signal at a time instant $t - \tau$ to $D$, where $\tau$ is the relay's processing delay. Without lost of generality, we model $x_j$ as an artificial Gaussian noise. In addition, we assume that D has a full knowledge of $x_j$ and global CSI. The first assumption can be achieved by using pseudo-random or chaotic sequences which are known to both $S$ and $D$, yet not open to $R$. The second assumption can be readily satisfied through the CSI exchange procedure before data transmission. With the aforementioned assumptions, $D$ can perform interference cancellation before decoding the desired signal. The SINRs at $R$ and $D$ (after interference cancellation) are given as follows
{\normalsize
\begin{align}
{\gamma _R} = \frac{{\alpha {\gamma _{sr}}}}{{\left( {1 - \alpha } \right){\gamma _{sr}} + {\gamma _{rr}} + 1}},
\end{align}
\begin{align}
{\gamma _D} = \frac{{\alpha {\gamma _{sr}}{\gamma _{rd}}}}{{\frac{{{\gamma _{rd}}{\gamma _{rr}}\left( {\alpha {\gamma _{sr}} + 1} \right)}}{{{\gamma _{sr}} + 1}} + {\gamma _{sr}} + {\gamma _{rd}} + {\gamma _{rr}} + 1}},
\end{align} }where ${\gamma _{sr}} = {P_s}{\left| {{h_{sr}}} \right|^2}/{N_0}$, $\gamma _{rd} = P_r \left| {{h_{rd}}} \right|^2 / N_0$, $\gamma _{rr} = {P_r}{\left| {{h_{rr}}} \right|^2}/{N_0}$, $P_r$ is the relay's transmit power, and $h_{sr}$, $h_rd$, and $h_rr$ denote the channel coefficients of the $S-R$, $R-D$, and the loop-interference channels, respectively. Noting that by setting $\alpha = 1$ in equations (1) and (2) we will obtain the SINRs at $R$ and $D$ with the conventional FDR scheme (without SBJ).
\begin{figure}[t]
  \centering
  \includegraphics[width=8cm]{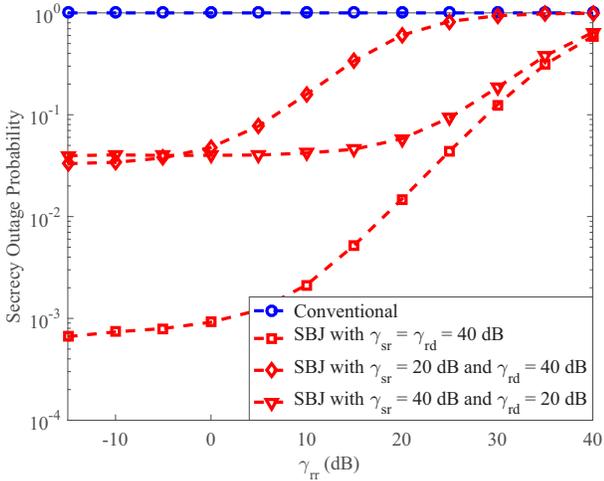}
  \caption{Secrecy outage probability versus $\gamma_{rr}$ of the conventional FD and the SBJ schemes.}
\end{figure}

We now present several representative simulation results to illustrate the characteristics of the proposed SBJ scheme. In the simulation, all the wireless channels are modeled as flat Rayleigh fading. In Fig. 6, we simulate the system SOP versus $\gamma_{rr}$ for the conventional FDR and the SBJ schemes with several values of $\gamma_{sr}$, $\gamma_{rd}$, and $\alpha = 0.5$. It is first shown that the SBJ scheme significantly outperforms the conventional FDR counterpart, whose SOP is always one. Secondly, for the SBJ scheme, decreasing the value of $\gamma_{sr}$ or $\gamma_{rd}$ increases the SOP since the received SINR at $D$ is upper bounded by the SINR of the weakest link. In addition, from the two middle curves, we observe that the SOP of the case $\gamma_{sr} = 20$ dB and $\gamma_{rd} = 40$ dB converges to one much faster than that of the case $\gamma_{sr} = 40$ dB and $\gamma_{rd}$ = 20 dB. It means that the $S-R$ channel has a greater impact on the SOP than the $R-D$ counterpart.
\begin{figure}[t]
  \centering
  \includegraphics[width=8cm]{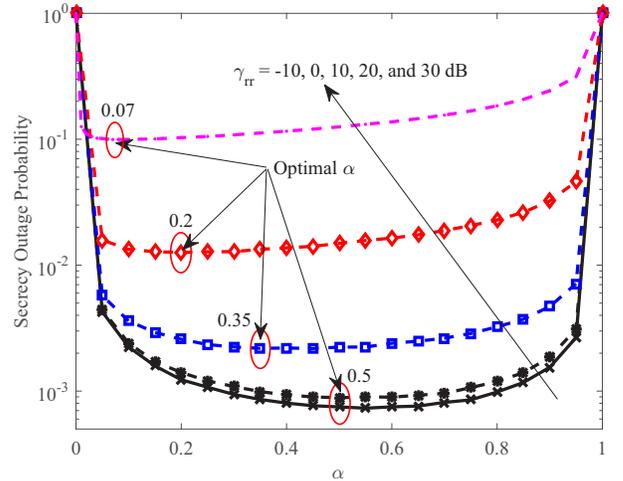}
  \caption{Secrecy outage probability versus $\alpha$ of the SBJ scheme with several values of $\gamma_{rr}$ and with $\gamma_{sr} = \gamma_{rd} = 40$ dB.}
\end{figure}

Figure 7 illustrates the SOP of the SBJ scheme versus the power allocation ratio $\alpha$. The figure shows that as $\gamma_{rr}$ increases, the SOP is enlarged. The reason is that the relay's LI is amplified and forwarded to the destination, and thus, the relay's LI would have a greater destructive impact on the received SINR obtained at $D$ than that obtained at $R$. In addition, it is seen that as $\gamma_{rr}$ is much smaller than $\gamma_{sr}$ and $\gamma_{rd}$, i.e. $\gamma_{rr} = -10$ and $0$ dB, it is optimal to equally allocate the power $P_s$ to the confidential and the jamming signals. Moreover, as $\gamma_{rr}$ increases, the optimal values of $\alpha$ deviates from 0.5 and tends to values that are close to zero. The explanation is as follows. As shown in the Fig. 6, when $\gamma_{rr}$ increases, the SOP of the SBJ scheme tends to one. It means that the SINR obtained at $D$ approaches to that obtained at $R$. Therefore, to enhance the system secrecy performance, we will need to allocate more power to the jamming signal $x_j$ to confuse the relay, and thus, reduces the relay's capacity.
\section{Challenges and Research Opportunities}
The scope of future research on this topic is broad, and we surely believe that novel cooperative scenarios and security schemes shall soon be proposed and developed. We now outline only a few interesting and challenging problems that are worth further consideration. First, the robustness of the existing K-PLS schemes to inexact CSI need to be investigated. It is obvious that the secrecy performance of the existing K-PLS schemes depends heavily on CSI. In reality, CSI should be first estimated and then fed back to a requesting node. However, due to estimation error and feedback delay, only imperfect CSI is available, which will definitely affect the K-PLS schemes' performance. Although CSI is a crucial factor for the K-PLS schemes, the effects of imperfect CSI on their secrecy performance have not been tackled yet.

Secondly, it is noted that all of the existing works focus only on trusted FD relays, but have ignored scenarios with untrustworthy FD relays. When relays are unstrusted, they are helpers and eavesdroppers simultaneously. To fully understand the true benefits of FD relays in enhancing security, cooperative systems with untrusted FD relays should be taken into account. For these systems, closed-form analytic expressions of the secrecy performance would be necessary to quickly predict the systems' behaviors in various environments and reveal possible novel insights about the effects of key parameters on the systems' behaviors, from which effectively parameters' tuning can be carried out. In addition, proposing an efficient power allocation scheme that can maximize the systems' secrecy performance under total or the individual power constraints is also a promising issue.

Thirdly, besides studying cases with all trusted or untrusted FD relays, considering heterogeneous FD systems, which consist of both trusted and untrusted relays, is also important. The reason is that heterogeneous systems would commonly exist in reality, i.e. in a sensor system, only a few relays are compromised and become adversaries, while the rest of relays remain trusted. For such systems, existing analytic performance results may not hold. In addition, existing K-PLS schemes may perform badly. Therefore, considering heterogeneous systems, deriving systems' secrecy performance, and proposing novel K-PLS schemes are necessary.

Fourthly, it is critical to search for new PLS techniques that can help cooperative systems defense against smart adversaries. A smart adversary can adaptively switch between eavesdropping and jamming based upon the quality of the wiretap channels. Particularly, if the wiretap channels are not good, eavesdropping may not work well, and thus, the smart adversary can instead send jamming signals to more efficiently disrupt ongoing communications \cite{Zou'16}. Under these scenarios, we should take into account both CSI of the interference links spanning from the adversary to the legitimate receiver and CSI of the wiretap channels during our design process.

Last but not the least, it is well-known that PLS can operate independent of higher layers so that it can be used to augment existing cryptographic schemes. However, up to now, research works that jointly consider PLS and cryptography have not been reported yet. In other words, cross-layer security approaches have not yet explored. Such cross-layer security approaches are expected to further improve the security level of cooperative systems at a lower cost as compared to conventional security mechanisms. Hence, designing cross-layer security schemes for cooperative systems is also an interesting problem.
\section{Conclusion}
In this work, we first presented a contemporary summary on K-PLS schemes proposed for wireless cooperative systems with FD relays. The focus was on schemes with both single and multiple relays to illustrate that FD relays can also be exploited to improve the secrecy performance of cooperative systems. By comparing the SOP of the existing K-PLS schemes, we observed that for single relay systems, the H-HD-FD scheme can be combined with the FDJ scheme to further enhance the systems' security in the low region of the residual LI. In addition, for systems with multiple relays, the Max-Min FD RS criterion, instead of the optimal FD RS counterpart, can be used in the H-HD-FD RS scheme to considerably reduce the systems' complexity without sacrificing much secrecy performance. Moreover, to improve the secrecy performance of a cooperative system with an untrusted FD relay, we proposed the SBJ scheme in which a source uses a fraction of its power to transmit jamming signal to confuse the untrustworthy relay. Finally, we provided an informative discussion on possible interesting, yet challenging, research problems that are worth further investigation.
\section*{Acknowledgment}
The authors gratefully acknowledge the support from Electronic Warfare Research Center at Gwangju Institute of Science and Technology (GIST), originally funded by Defense Acquisition Program Administration (DAPA) and Agency for Defense Development (ADD).

\end{document}